\documentstyle[12pt,epsf,epsfig]{article}

\begin{document}
\begin{titlepage}
\pagestyle{empty}
\baselineskip=14pt
\rightline{UMN-TH-1704/98}
\rightline{LBL-41768}
\rightline{UCB-PTH-98/23}
\rightline{hep-ph/9805300}
\hfill \today
\vskip .2in
\baselineskip=18pt
\begin{center}
{\large{\bf Regulating the Baryon Asymmetry in No-Scale \\
 Affleck-Dine Baryogenesis}
}\footnote{This work was supported in part
by DoE grants
DE-FG02-94ER-40823 and DE-AC03-76SF00098, and by NSF grants  
AST-91-20005
and PHY-95-14797 and by the Alfred P. Sloan Foundation.}
\end{center}  
\vskip .1in
\begin{center}

Bruce A. Campbell  

{\it Department of Physics, University of Alberta}

{\it Edmonton, Alberta, Canada T6G 2J1}

Mary K. Gaillard and Hitoshi Murayama   

{\it Department of Physics and Theoretical Physics Group, Lawrence  
Berkeley Laboratory}

{\it University of California, Berkeley, California, 94720}

 and

Keith A. Olive

{\it School of Physics and Astronomy, University of Minnesota}

{\it Minneapolis, MN 55455, USA}

\vskip .1in
\end{center}
\vskip .2in
\centerline{ {\bf Abstract} }
\baselineskip=18pt
In supergravity models (such as standard superstring  constructions) that
possess a Heisenberg symmetry, supersymmetry breaking by the inflationary
vacuum energy does not lift  flat directions at tree level. One-loop
corrections give small squared masses  that are negative ($\sim - g^2
H^2/(4\pi)^2$) for all flat   directions that do not involve the stop. After
inflation, these flat  directions generate a large baryon asymmetry;
typically $n_B/s \sim$ O(1).   We consider mechanisms for suppressing this
asymmetry to the observed level. These include dilution from inflaton or
moduli decay,  GUT nonflatness of the $vev$ direction, and higher
dimensional   operators in both GUT models and the MSSM.  
We find that the observed BAU can easily be generated when one or more 
of these effects is present.

\noindent
\end{titlepage}
\renewcommand{\thepage}{\roman{page}}
\setcounter{page}{2}
\mbox{ }

\vskip 1in

\begin{center}
{\bf Disclaimer}
\end{center}

\vskip .2in

\begin{scriptsize}
\begin{quotation}
This document was prepared as an account of work sponsored by the United
States Government. While this document is believed to contain correct 
information, neither the United States Government nor any agency
thereof, nor The Regents of the University of California, nor any of their
employees, makes any warranty, express or implied, or assumes any legal
liability or responsibility for the accuracy, completeness, or usefulness
of any information, apparatus, product, or process disclosed, or represents
that its use would not infringe privately owned rights.  Reference herein
to any specific commercial products process, or service by its trade name,
trademark, manufacturer, or otherwise, does not necessarily constitute or
imply its endorsement, recommendation, or favoring by the United States
Government or any agency thereof, or The Regents of the University of
California.  The views and opinions of authors expressed herein do not
necessarily state or reflect those of the United States Government or any
agency thereof, or The Regents of the University of California.
\end{quotation}
\end{scriptsize}

\vskip 2in

\begin{center}
\begin{small}
{\it Lawrence Berkeley Laboratory is an equal opportunity employer.}
\end{small}
\end{center}

\newpage
\renewcommand{\thepage}{\arabic{page}}
\setcounter{page}{1}
\baselineskip=18pt
 
\def\({\left(}
\def\){\right)}
\def\[{\left[}
\def\]{\right]}
\def\la{~\mbox{\raisebox{-.6ex}{$\stackrel{<}{\sim}$}}~}
\def\ga{~\mbox{\raisebox{-.6ex}{$\stackrel{>}{\sim}$}}~}
\def\b-l{$B - L$}
\def\bl{$B + L$~}
\def\tm{$\tilde m$}
\def\beq{\begin{equation}}
\def\eeq{\end{equation}}
\def\bea{\begin{eqnarray}}
\def\eea{\end{eqnarray}}

\topmargin -0.5in
\textheight 8in
\setcounter{footnote}{0}

\section{Introduction}

Supersymmetric (SUSY) theories possess a unique and efficient mechanism for the
generation of a cosmological baryon asymmetry through the decay of scalar
fields along nearly $F$- and $D$-flat directions of the scalar  
potential, using baryon number violation arising in GUTS \cite{ad}, 
or other sources of baryon or lepton number violation such as neutrino 
mass effects \cite{cdo1,cdo2,my}.  In its original form, 
applied to SUSY-GUTS, this Affleck-Dine (AD) mechanism  not only
produces a baryon asymmetry but also entropy from the decaying fields  
and results in a net baryon-to-entropy ratio, $n_B/s \sim$ O(1)
\cite{ad,lin}. In fact, one of the main problems associated with this  
mechanism in the GUT case
is the dilution of the asymmetry down to acceptable levels of order
$10^{-11}$--$10^{-10}$ \cite{eno}.
In the context of GUTS after inflation, the AD scenario
can be shown \cite{eeno} to produce an asymmetry of the desired  
magnitude. 
During inflation, scalar fields with masses,
$m < H$, are driven  by quantum fluctuations
to large vacuum values ( but less than $M_P$) which become the source of the
scalar field   oscillations once inflation is over: however, inflation
supplies  an additional source of entropy from inflaton decays. 

Recently, it has been argued that the simple picture of driving  
scalar fields to large vacuum values along flat directions during inflation is  
dramatically altered
in the context of supergravity \cite{drt}.  During inflation, the  
Universe is dominated by the vacuum energy density, $V \sim H^2 M_P^2$. The  
presence of a nonvanishing and positive vacuum energy density indicates that  
supersymmetry is broken and soft masses of order of
$\sim H$ are generated \cite{all}. If such masses are generated along
otherwise flat directions, the AD mechanism would be inoperative.
While this is true for minimal supergravity, it was shown \cite{gmo}
that in no-scale supergravity \cite{ns}, or more generally in any  
supergravity
theory with a Heisenberg symmetry of the kinetic function \cite{heis},
such corrections are absent at the tree level.  It was also  
shown
\cite{gmo} that, at the one-loop level, corrections to the scalar  
potential
are generated along flat directions and, for all directions which do  
not involve
stops, these corrections to the mass squared are negative of  
magnitude 
$\sim 10^{-2} H^2$.  This is large enough to insure that the $vev$'s  
along
flat directions do indeed run to $\phi_0 \sim M_P$, resulting once  
more in a
baryon asymmetry of order unity.
 
In this letter, we will systematically consider possible effects  
which may
contribute to diluting a baryon asymmetry $n_B/s \sim$ O(1) from the  
AD
mechanism.  We will consider in turn the effects of dilution by an  
inflaton
and/or a massive moduli field, the effects of higher dimensional  
operators, and 
the effects of an intermediate scale.  We find that the dilution by  
an inflaton or moduli is insufficient in generic supergravity models, 
but may be sufficient in the context of certain string-derived models.
In the absence of these effects
the presence of the first nonrenormalizable contribution to the  
superpotential (of dimension 4), suppressed by the GUT scale, can lead  
to the desired asymmetry.

We also consider baryogenesis in the context of the minimal supersymmetric
standard model, with only the baryon number violation, and nonperturbative
corrections to the superpotential, implied by quantum gravity effects at
the Planck scale. We find that these effects are sufficient to generate the 
observed baryon asymmetry of the universe (BAU) in the case of the minimal
supersymmetric standard model with standard 4-dimensional (super)gravity. 
We also consider the possibility that the asymmetry is diluted by sphaleron
effects at the electroweak scale.


\section{Generalities}

 The contribution to scalar
 masses can be derived from the scalar potential in a supergravity model
described
 by a K\"ahler potential $G$ \cite{sugr},
 \beq  V = e^G \left[ G_i {(G^{-1})}^i_j G^j - 3 \right] \label{gen}  \eeq
 where $G_i = \partial G/\partial \phi^i$ and $G^i =
 \partial G/\partial \phi^*_i$ and we are setting the D-terms equal to 0,
which is true for gauge singlets and along the $D$-flat directions.
In minimal supergravity, the  K\"ahler potential is defined by
 \beq  G = zz^* + \phi_i^* \phi^i + \ln |\overline{W}(z) + W(\phi)|^2
 \label{min}  \eeq
where $z$ is a Polonyi-like field \cite{pol} needed to break supersymmetry, and
we denote the scalar components of the usual matter chiral
supermultiplets by $\phi^i$. $W$ and $\overline{W}$
are the superpotentials of $\phi^i$ and $z$ respectively. In this case,
the scalar potential becomes
 \beq
 V = e^{zz^* + \phi_i^* \phi^i} \left[ |\overline{W}_z +
 z^* (\overline{W} + W)|^2
      + |W_{\phi^i} + \phi^*_i (\overline{W} + W)|^2 -  
3|(\overline{W} + W)|^2
    \right] .
 \eeq
 In the above expression for $V$, one finds a mass term for  
the matter
 fields $\phi^i$, $e^G  \phi_i^* \phi^i = m_{3/2}^2 \phi_i^* \phi^i$
 \cite{bfs}.
Scalar squared masses
 will pick up a contribution of order $m_{3/2}^2 \sim V/M_P^2 \sim  
H^2$,
 thus lifting the flat directions and potentially preventing the   
realization of
 the AD scenario as argued in \cite{drt}.  Below we use
 reduced Planck units: $M_P = 1/\sqrt{8\pi G_N} = 1$.
All flat directions in the matter sector are  
lifted as well.

In models with a  ``Heisenberg symmetry" \cite{heis}
the dangerous cross-term is absent. If in addition to the field $z$  
and matter
fields $\phi^i$, we allow for hidden sector fields $y^i$, the 
``Heisenberg symmetry" is defined by
\begin{equation}
\delta z = \epsilon^*_i \phi^i, \hspace{1cm} \delta \phi^i =  
\epsilon^i ,
\hspace{1cm} \delta y^a = 0, \label{heis}
\end{equation}
where $\epsilon^i$ are complex parameters and $\epsilon_i^*$ their
complex conjugates. 
The combination
\begin{equation}
\eta \equiv z + z^* - \phi_i^* \phi^i . \label{etadef}
\end{equation}
and $y_i$ are invariant.
Let us assume this is a symmetry of the kinetic function in the  
K\"ahler
potential.  We also require that the field $z$ does not have a  
coupling
in the superpotential. 
Then the most general K\"ahler potential becomes
\begin{equation}
G = f(\eta) + \ln |W(\phi)|^2 + g(y), \label{kpot}
\end{equation}
where the superpotential $W$ is a holomorphic function of $\phi^i$  
only, and the fields $\eta$ and $\phi^i$ are regarded as independent degrees  
of freedom \cite{Gel,bg,gmo}, yielding a potential \cite{MSYY2}:
\begin{equation}
V = e^{f(\eta)+g(y)} \left[
	\left( \frac{f^{\prime 2}}{f^{\prime\prime}} - 3 \right)
		|W|^2
	- \frac{1}{f'} |W_i|^2 + g_a(g^{-1})^a_bg^b |W|^2\right] .
\label{hpot}
\end{equation}
It is important to notice that the cross term $|\phi_i^* W|^2$ has 
disappeared in the
scalar potential.  Because of the absence of the cross term, flat
directions remain flat even during inflation.  
A detailed discussion of this result is found in \cite{gmo}.

The above result however, is only valid at the tree-level.
Since gravitational interactions preserve the
Heisenberg symmetry at one-loop \cite{heis}, the only possible
contribution to the mass of the flat directions can come from either  
gauge
interactions or superpotential couplings that contribute to the  
renormalized
K\"ahler potential.  This too was computed in \cite{gmo} from the  
general
results~\cite{tr,TR} for the  one-loop corrected supergravity  
Lagrangian.
 Assuming that inflation is driven by an $F$-term
rather than a $D$-term (the result is similar if $\langle D\rangle\ne  
0$), the
one-loop corrected mass is
\begin{eqnarray}
\left( m^2 \right)_i^j &=&  
\frac{\ln(\Lambda^2/\mu^2)}{32\pi^2}\Bigg\{ h_{ikl} 
h^{*jkl} \left[\alpha\langle V \rangle + m_{3/2}^2\left(5{f'^2\over  
f''} + 
2{f'f'''\over f''^2} - 10 - {f'^4\over f''^2}\right)
 \right] \nonumber \\ & &
- 4\delta^j_ig_a^2C_2^a(R_i)\left[\beta\langle V \rangle  
+ m_{3/2}^2\left({f'f'''\over f''^2} - 2\right) \right]\Bigg\} ,
\label{mass}\end{eqnarray} 
where the vacuum energy $\langle V \rangle $ and the gravitino mass  
$m_{3/2}$
are their values during inflation.
 We are assuming that the inflaton, $\psi$, is one of the
$\phi^i$ or a hidden sector $y^i$ field.  
 $\mu\ge \langle V\rangle $ is the 
appropriate infrared cut-off in the loop integral, and 
$\Lambda$ is the cutoff scale below which the effective supergravity
Lagrangian given by the K\"ahler potential eq.~(6) is valid; $\Lambda  
= 1$ in many models but can be lower if the effective potential
for inflation is generated at an intermediate scale. 
The $h$'s are Yukawa coupling constants
$g_a,C_2^a(R_i)$ are the coupling constant and matter Casimir for the 
factor gauge group $G_a$, and the parameter $\alpha,\beta,$ are model 
dependent.  In the no-scale case $f = - 3\ln\eta$, the result
(\ref{mass}) reduces to
\begin{equation}
\left( m^2 \right)_i^j = \frac{\ln(\Lambda^2/\mu^2)}{32\pi^2}\langle  
V \rangle 
\left[\alpha h_{ikl} h^{*jkl} - 4\beta\delta^j_ig_a^2C_2^a(R_i)  
\right].
\label{nsmass}\end{equation}
If the inflaton is one of the $\phi^i$ ($\phi_0\ne 0$), $\alpha =  
\beta =
{2\over3}$. If the
inflaton is in the hidden sector ($y_a\ne 0$), $\alpha < 0,\;
\beta = 1$ if the inflaton is not the dilaton $s: \;\langle s\rangle  
= g^{-2}$.  
If the inflaton is the standard string dilaton, $\beta = \alpha =  
-1$.
In all but the last case the masses are negative if gauge couplings  
dominate
Yukawa couplings.  We will assume that the inflaton is not the  
dilaton.

The vacuum energy in Eq. (\ref{nsmass}) is related to the expansion  
rate by
$H^2 = (1/3) \langle V \rangle$, and hence the typical mass of the  
flat
directions is $m^2 \simeq 10^{-2} H^2$ during inflation. However,  
from
Eq. (\ref{nsmass}), we see that for all scalar matter fields aside  
from stops,
the contribution to the mass squared is {\em negative} as the Yukawa  
couplings
are smaller than the gauge couplings.  Thus any flat direction not  
involving
stops will have a negative contribution at one-loop without an ad hoc  
choice of
the parameters. Now, as argued in \cite{gmo}, even though  
fluctuations
will begin the growth of $\phi_0$, the classical equations of motion
soon take over.  The classical equations of motion drive $\phi_0$ as
$(- m^2) t$ which is smaller than the quantum growth only for 
$Ht < H^4/m^4$.  Then for $Ht > H^2/(- m^2)$, the classical growth of
$\phi_0$, becomes nonlinear $\sim H e^{-m^2 t/3H}$, 
and $\phi_0$ will run off to its minimum
determined by the one-loop corrections to $\phi^4$, which are again
of order $V$.  An explicit one-loop calculation~\cite{tr,TR} shows  
that the
effective potential along the flat direction has the form 
\begin{equation}
V_{eff} \sim \frac{g^2}{(4\pi)^2}
\langle V \rangle \left( 
	-2 \phi^2 \log \left(\frac{\Lambda^2}{g^2 \phi^2}\right) 
	+ \phi^2 \right)
	+ {\cal O}(\langle V \rangle)^2 ,
\label{finalV} \end{equation}
where $\Lambda$ is the cutoff of the effective supergravity theory,  
and
has a minimum around $\phi \simeq 0.5 \Lambda$.  This is 
consistent because this effective
potential is only of order $- \langle V \rangle g^2/(4\pi)^2$ and is  
a
small correction to the inflaton energy density which drives  
inflation. 
Thus, $\phi_0 \sim M_P$ will be generated and 
in this case the subsequent sfermion oscillations will
dominate the energy density and a baryon asymmetry will result 
which is independent of inflationary parameters as originally  
discussed in
\cite{ad,lin} and will produce $n_B/s \sim O(1)$.

For  a $\phi_0$ $vev$ as large as $M_P$, coherent oscillations of the  
flat
directions will persist long after the inflatons have decayed and the  
radiation
produced in their decays has redshifted away.
To see this \cite{eeno}, consider the energy density stored in the  
inflaton and
the scalar fields making up our flat direction
\begin{eqnarray}
\rho_\psi & = & m_\psi^2 \psi^2 \\
\rho_\phi & = & m_\phi^2 \phi^2
\end{eqnarray}
We will assume that the initial $vev$ for $\psi$ is $M_P$
and the initial $vev$ for $\phi$ is $\phi_0$, which as we saw above,  
due to the
one-loop correction to the flat direction is also $M_P$.
After inflation (the de Sitter expansion), the corrections in Eq.  
(\ref{finalV})
are turned off and the flat direction is restored up to supersymmetry  
breaking
effects of order ${\tilde m} \sim 10^{-16}$. 
If we consider only  a generic model of inflation with one single  
input scale
determined by the COBE normalization of the microwave background  
anisotropy,
then $m_\psi \sim 10^{-7}$ \cite{eeno,cdo2} and the inflatons begin  
oscillating
about the minimum of the inflationary potential when the Hubble  
parameter $H
\simeq m_\psi \psi/M_P \sim m_\psi$.  Let us denote the value of the  
scale
factor $R$ at this point $R_\psi$.  For $R > R_\psi$, the Universe  
expands as
if it were matter dominated so that $H \simeq m_\psi  
(R_\psi/R)^{3/2}$.
The two field system involving the flat direction and the inflaton  
was worked
out in detail in \cite{eeno}.  We only quote the relevant results  
here.
Inflatons decay at $R_{d\psi} \simeq (M_P/m_\psi)^{4/3} R_\psi$ when  
$\Gamma_\psi =  m_\psi^3/M_P^2 \simeq H$ (their
decay rate is assumed to be gravitational). 
After inflaton decay, the Universe is radiation dominated and  
$\rho_{r\psi}
\simeq m_\psi^{2/3} M_P^{10/3} (R_\psi/R)^4$, where $\rho_{r\psi}$ is  
the
energy density in the radiation produced by inflaton decay.

Similarly, the fields $\phi$ begin oscillating about the origin, when
$H \sim {\tilde m} > \Gamma_{\psi}$.  After $\phi$'s begin oscillating at $R =  
R_\phi$, their
energy density  is just $\rho_\phi \simeq {\tilde m}^2 \phi_0^2  
(R_\phi/R)^3
\simeq m_\psi^2 \phi_0^2 (R_\psi/R)^3$.  In the AD mechanism, one 
utilizes the flat directions lifted only by supersymmetry breaking 
effects
\begin{equation}
	\int d^{4}\theta \zeta^{*} \zeta \frac{\phi^{*} \phi^{*} \phi 
	\phi}{M_{X}^{2}} = \tilde{m}^{2} \frac{\phi^{*} \phi^{*} \phi 
	\phi}{M_{X}^{2}},
\end{equation}
where $\zeta = \tilde{m} \theta^{2}$ is the supersymmetry-breaking 
spurion.  Such operators can carry non-vanishing baryon-numbers.
The flat  directions
associated with baryon number violating operators store a net baryon  
number
density $n_B = \epsilon (\phi_0^2/M_X^2) \rho_\phi/{\tilde m}$, where  
$M_X$
is the scale of the baryon number violating operator, $M_X \sim  
10^{-3}$ for
GUT scale baryon number violation and $\epsilon$ is measure of CP  
violation in
the oscillations and should be O(1). Oscillations along the flat  
direction
decay when their decay rate
$\Gamma_\phi \simeq {\tilde m}^3 / \phi^2 \sim H$. 
If 
\beq
\phi_0 > {\tilde m}^{5/12} M_P^{4/3} / m_\psi^{3/4} \sim  
10^{-3/2}
\label{cond}
\eeq
the
flat direction fields decay so late that the energy density in  
radiation from
inflaton decays,
$\rho_{r\psi}$ has redshifted away and the Universe is dominated by  
$\phi$'s
when they decay.  In this case, the resulting entropy in the Universe  
is
produced by $\phi$ decay and $n_B/s \sim $ O(1) \cite{ad,lin}.
(In fact, if the asymmetry were computed solely from sfermion decays as in
\cite{ad}, it might appear that a baryon to entropy ratio greater than 1,
$n_B/s \sim (M_P/{\tilde m})^{1/6}$, were possible.  However as argued by
Linde
\cite{lin}, the maximum asymmetry achieved  is $\sim 1$, independent of the
the initial value of $\phi_0$.) For $\phi_0 < {\tilde m}^{5/12} M_P^{4/3} /
m_\psi^{3/4}$,
$\phi$'s   decay while
the Universe is dominated by
$\rho_{r\psi}$. The Universe will remain radiation dominated and the
baryon-to entropy ratio is   that
computed in \cite{eeno}
\beq
\frac{n_B}{s} \sim  \frac{\epsilon {\phi_0}^4 {m_\psi}^{3/2}}
{({M_X}^2 + \phi_0^2) {M_P}^{5/2} \tilde{m}} 
  \la  \min (1, 10^{6} ({\phi_0 \over M_P})^2, 10^{12} ({\phi_0 \over
M_P})^4) .
\label{eenonb}
\eeq
In eq. (\ref{eenonb}), when the condition (\ref{cond}) is satisfied, the 
asymmetry is O(1). For smaller $\phi_0$, the asymmetry is given by one
of the two expressions on the right of (\ref{eenonb}), depending on whether
$\phi_0$ is greater or less than $M_X$.   
Finally, for $\phi_0 < {\tilde m}^{5/2} M_P^{3} / m_\psi^{9/2}$, the  
flat
direction fields decay before the inflatons, but in this case, too  
small an
asymmetry would result.   These results are distinguished along the  
right axes
in the Figure. Values of $\tilde m$ and $M_X$ were taken to be $10^{-16}$
and $10^{-3}$ respectively in eq. (\ref{eenonb}) and in the Figure. In the
absence of the types of effects we discuss   below, we would  expect $\phi_0
\sim $ O(1) due to the one-loop potential   (\ref{finalV})
and our baryon asymmetry is too large.  This is the source of our  
problem.
 
\section{Dilution by Moduli Decay}

Polonyi-like  moduli (which we will label as $y$) have been known to  
be a source
of embarrassment primarily due to their excessive entropy production  
\cite{pp}.
In a conventional GUT baryogenesis scenario, the late decay of moduli  
would lead
to an entropy dilution factor of O($10^{16}$), and lead to a low  
reheat
temperature of about 1 keV unless $m_y \ga 10$ TeV, similar to the  
cosmological
gravitino problem \cite{sw}. Furthermore, in models with $R$-parity
conservation, it was pointed out that unless $m_y \ga 10^6$ GeV, the  
decay of
$y$ oscillations will lead to an unacceptably large LSP relic density
\cite{kmy}. This is the shaded region on the left in the figure.  Because
of the potential problem with entropy   production, the
$y$'s seem to be a natural place to look for the dilution of the AD  
baryon asymmetry. 

For now, we will treat the moduli mass as a free parameter, $m_y$.
Unless, $m_y < m_\psi$, the moduli would dominate the energy density  
of the
Universe early on ($\rho_y \simeq m_y^2 y_0^2$ and $y_0 \sim M_P$)  
contrary to
our assumption that inflation is driven by the inflaton.  For $m_y <  
m_\psi$,
the moduli oscillations live on after inflaton decay, and can  
subsequently come
to dominate the energy density of the Universe.  As in the  
$\phi$-$\psi$
system, when $\phi_0 > {\tilde m}^{5/12} M_P^{4/3} / m_y^{3/4}$,  
$\phi$'s come
to dominate and a baryon-to-entropy ratio of O(1)  is produced as  
displayed in
the upper right corner of the figure (above the solid grey line).  For smaller
$\phi_0$, the produced asymmetry will be given by
\beq
\frac{n_B}{s} \sim  \frac{\epsilon {\phi_0}^4 {m_y}^{3/2}}
{({M_X}^2 + \phi_0^2) {M_P}^{5/2} \tilde{m}} 
\label{ynb}
\eeq
and when $\phi_0 < {\tilde m}^{5/2} M_P^{3} / m_y^{9/2}$, $y$'s decay  
after
$\phi$'s.  This region is demarked by the dashed line in the Figure.
Thus at $\phi = M_P$, this is the only region relevant for the  
dilution of the baryon asymmetry.  For Planck scale baryon number
violation of the type discussed below, we show the region in the 
$m_y-\phi_0$ plane which results in a baryon asymmetry of $10^{-10} -
10^{-11}$. Clearly a $vev$ of order O(1) for $\phi$ results in too large
an asymmetry no matter what value of $m_y$ we choose.  GUT scale 
baryon number violation would only compound this problem.


If we implement the constraint on $m_y$ from their decays into LSP's
\cite{kmy}, and consider $m_y \simeq 10^6$ GeV, and $\phi_0 = M_P$,  
Eq.
(\ref{ynb}) gives $n_B/s \simeq 10^{-7/2}$, and we see that the  
dilution is far
too small.  At $m_y = 10$ TeV, $n_B/s \simeq 10^{-13/2}$, the  
dilution is still
insufficient.  Even at $m_y = 100$ GeV, $n_B/s \simeq 10^{-19/2}$,  
the asymmetry is somewhat too large. As one can see, the
primary moduli   problem of
entropy production is nonexistent in the AD baryogenesis scenario.
Whether or not moduli can indeed be heavy enough to avoid a low  
reheating temperature or LSP production is another issue which has also been  
considered
in the context of no-scale supergravity \cite{bg,enq} and recent  
string theory formulations \cite{bd,bgw3,bcc}. 

If however R-parity is violated the constraint from an over-density
in LSPs can be avoided and smaller values of $m_y$ are allowed.
For example, in orbifold compactifications of the heterotic string, 
the untwisted sector of the effective field theory is Heisenberg invariant at
the classical level.
It has recently been shown, both in a modular invariant linear multiplet
formulation~\cite{bgw1} and in a chiral multiplet formulation~\cite{cas} for the
dilaton supermultiplet, that dilaton stabilization with string-scale weak 
coupling ($g_{st}\sim 1$) can be achieved by invoking string nonperturbative
effects~\cite{shenk}.  It has further been shown~\cite{bgw3,bcc} that in these 
models the moduli problem of low temperature reheating can be avoided ($m_y > 
10$~TeV with $y$ a modulus $t_I$ or the dilaton).  Here we consider the 
quasi-realistic, modular
invariant, model of~\cite{bgw2,bgw3} with both gaugino and matter condensation, 
and moduli stabilization at a self-dual point $t_I=1$ or $e^{i\pi/6}$ in the 
true vacuum. The condensation scale is
\beq \Lambda_c \sim e^{-1/6b(2/g^2_{st}+\pi b_8)},\eeq 
where $3b$ is the $\beta$-function coefficient of
the strong hidden sector gauge group, and $b_8 = .38\gg b$ is the 
$E_8\;\beta$-function coefficient.  The masses of gravitino 
($m_{3\over2}$), moduli ($m_{t}$) and dilaton ($m_{d}$) in vacuum
are given by
\beq m_{3\over2} = {1\over4}b\Lambda^3_c \approx {b\over 2b_8}m_t \approx
b^2 m_d. \eeq
Thus $m_d>m_t$, and the moduli reheating problem is resolved provided $b\ge
.016$--.027 for $g_{st}= 1$--.5, in which case $m_{3\over2} \ge .74$--1.3~TeV
and $\Lambda_c\ge .9\times 10^{14}$~GeV.
Scalar masses depend on their as yet unknown couplings to the Green-Schwarz term
needed to restore modular invariance.  If they are decoupled from this term,
their masses, as specified at the condensation scale, are equal to the 
gravitino mass.  A plausible alternative is that
the untwisted sector fields couple to the GS term in the
same Heisenberg invariant
combination that appears in the untwisted sector K\"ahler potential, {\it i.e.}
that the GS term depends only on the radii of compactification of the three 
tori. In this case the untwisted  sector masses are $m_u(\Lambda_c) = m_t/2$. 
(Multi-TeV masses for some squarks and sleptons have also been proposed in 
other contexts~\cite{cohen}, which are, however, subject to the 
stringent constraint from positivity of the scalar top mass squared \cite{Nima}.)
If we assume that ${\tilde m} \sim m_y \sim 10$~TeV in (14), where now 
${\tilde m}$ is the true vacuum mass of the particles along the relevant flat 
directions during inflation, one can obtain the observed baryon density of
$10^{-11}$--$10^{-10}$ with values of $\phi_0\sim 10^{-1}M_P$ 
if we identify $M_X$ 
with the string scale $\Lambda_{st} \sim g_{st}M_P$, as would be the case in an
affine level one model with no GUT below the string scale.  Since now 
$\phi_0< {\tilde m}^{5\over2}M^3_P/m^{9\over2}$, $y$'s decay after the
$\phi$'s and since $m_y \la 10^6$ GeV, the model falls in the shaded region
of the Figure, and its viability as a mechanism for diluting baryon number
requires R-parity violation.  
The required value of $\phi_0$ can be read off the $n_B/s$ contours
in the figure at low $m_y$ ($m_y/m_\psi \sim 10^{-8}$ in this case)
after accounting for the slightly different values of $M_X$ and $\tilde m$
chosen.

If R-parity is conserved, the LSP problem can be evaded only if the moduli
are stabilized at or near their ground state values during inflation.  In fact,
the domain of attraction near $t_1 = 1$ is rather limited: $0.6 < {\rm Re}t_I
<1.6$, and the entropy produced by dilaton decay with an initial value in this
range might be less that commonly assumed.  Assuming that the entropy
generated by the decay of the moduli $t$ is negligible, the entropy generated
by dilaton decay is not a problem in this model because the dilaton mass is
about $10^3$~TeV~\cite{bgw3}.  With untwisted sector masses $\sim 10$~TeV, we need
$\phi_0 \le 10^{-2}M_P$ if we include dilution from dilaton decay.  Thus the
amount of additional suppression needed is rather mild in these models.

One generally expects untwisted sector flat directions to be lifted by 
Heisenberg noninvariant terms arising from effects such as string loop threshold
corrections and nonperturbative corrections, that can contain factors of the
Dedekind function $\eta(t_I).$  For example, in the inflationary model
of~\cite{glm}, some of the moduli are stabilized at $t_I = e^{i\pi/6}$;
the flat directions in the corresponding untwisted sector are lifted: 
$m_{\phi^{AI}}\approx m_{t_I}\sim V^{1\over2}$, and cannot contribute to the
Affleck-Dine mechanism.   The potential for the 
remaining moduli remains flat up to corrections from the condensation 
potential:  $m_t \sim V_c^{1\over2} \sim 100$~TeV, and the flat directions 
$\phi$ in the corresponding untwisted sector have squared masses of the same 
order provided $|\phi_{IA}|^2\le {\rm t_I}$. This contribution is 
negative if $|\phi_{IA}|^2\le .2{\rm t_I}$, and is much 
smaller than that induced by the loop effects ($-m^2_\phi \sim 10^{-2}V$) in 
(10), and therefore cannot suppress the baryon number.   

\section{Regulation by Non-renormalizable Operators}

Hereafter we ignore possible effects from the decay of moduli (we do however
retain the dilution from inflaton decay, which is fixed and model  
independent provided that the inflaton decay is gravitational). 
We next consider the reduction in baryon asymmetry which results from the  
limitations
on the flat direction $vev$'s which are implied by the presence in the  
flat
direction potential of higher dimensional operators induced at the  
Planck scale
by gravitational dynamics, or at a GUT scale (should one exist) by  
the presence
of GUT interactions.

First we consider higher dimensional operators that are generically
expected in supergravity and superstring theory; these can limit cosmological 
$vev$'s that are generated along flat directions of the supersymmetric  
standard model (including the case where the standard model is embedded in a 
GUT with the flat direction in question continuing to remain flat in the  
embedding GUT). In general, the renormalizable terms in the superpotential 
of the supersymmetric standard model merely represent the lowest dimensional  
terms in an infinite series of ascending dimension; these leading terms define 
the low energy
Wilson effective action governing the dynamics of the standard model
superfields.  One expects that all terms of a  
given dimension that are not forbidden by gauge invariance will arise in  
the full effective action (global symmetries are generally violated by  
nonperturbative 
gravitational effects, and hence cannot prevent the presence of  
gravitationally induced terms in the effective superpotential).

On dimensional grounds one then expects that there will be a 
series of higher dimensional contributions to
the superpotential for the flat direction field $\Phi$, by terms of  
the form: $h\Phi^n/M_{P}^{n-3}$. These induce flat direction  
potential terms of the form:
$h^2 \phi^{(2n-2)} /M_{P}^{2n-6}$, where $h$ represents a dimensionless
coupling that incorporates the underlying dynamics, and we have exhibited the
gravitational scale dependence explicitly. As noted above, during inflation
there is a vacuum energy induced contribution to the potential of the flat
directions of the form  $\sim - g^2 H^2 \phi^2/(4\pi)^2$. Minimizing the
$vev$   potential, including higher dimensional corrections from a leading  
superpotential term of
order $n$ in $\Phi$ as above, we find that during inflation the the flat 
direction $vev$ rolls to a value given by: 
\beq
\phi^{(2n-4)} \sim \frac{2g^{2}}{(2n-2)h^2} \frac{H^2}{{(4\pi)}^{2}}
{M_{P}^{(2n-6)}}   .
\eeq
Dropping the numerical prefactor which involves the undetermined, but
generically comparable, couplings $g$ and $h$, and recalling that for  
viable inflationary models ${H}/{(4\pi)}$ is of order $10^{-8} M_{P}$, we  
have for the flat direction $vev$'s at the end of inflation: 
\beq \phi^{(2n-4)} \sim {(10^{-16})}{M_{P}^{(2n-4)}}  
\label{cut} \eeq
In particular, for the potential arising from the leading higher 
dimensional gravitationally induced
superpotential terms of dimension 6 ($n=4$), we have $\phi \sim
{10^{-4}}{M_{P}}$.  

We remark that higher dimension terms constructed from untwisted
sector fields in orbifold string compactifications necessarily contain factors
of the the Dedekind functions $\eta(t_I)$ that are needed to preserve modular
invariance, but break Heisenberg invariance.  If the superpotential contains a 
term $W(t) = \eta(t)^nw$, there is a corresponding contribution to the
potential: 
\beq V(t)\ni e^K|W(t)|^2(t + \bar{t})\[n^2(t + \bar{t})|\xi(t)|^2 -
n\(\xi(t) + {\rm h.c.}\)\],\quad \xi = {1\over\eta}{\partial\eta\over\partial t}
. \eeq
Such terms have been encountered in modular invariant parameterizations of 
gaugino condensation~\cite{eta} with $n=-6$; in that case they destabilize the 
potential 
in the direction of weak coupling (${\rm Re}s\to 0,\; K\ni -\ln(2{\rm Re}s)$)
because $\xi({\rm Re}t\ge 1)<0$, and $V(t)<0$ for $0\le t< 1.9$.  In the present
case, operators of higher dimension in untwisted fields require $n>0$, so $V(t)
\ge 0$ for Re$t\ge 1$ (it is sufficient to consider this domain because of
modular invariance of the potential), and such terms are not {\it a priori}
incompatible with a bounded potential.

As well as providing superpotential contributions to the flat direction
potential, the nonperturbative gravitational effects will induce higher
dimensional operators exhibiting baryon and lepton number violation, again
scaled by the appropriate powers of the Planck mass. In particular, the
resulting baryon and lepton number violation will, after  
supersymmetry breaking,
give rise to quartic scalar terms in the flat direction effective potential
which are now scaled by ${{\tilde{m}}^{2}}/{M_{P}^2}$. This results, after flat
direction oscillation and decay, in a baryon asymmetry from Eq.
(\ref{eenonb}) with the replacement $M_X \rightarrow M_P$
\beq
\frac{n_B}{s} \sim  \frac{\epsilon {\phi_0}^4 {m_\psi}^{3/2}}
{ {M_P}^{9/2} \tilde{m}} 
  \sim  ~ \epsilon ~~ 10^{6} ({\phi_0 \over M_P})^4
\label{eenonc}
\eeq
The baryon asymmetry in this case is shown on the right hand side of the
Figure ($m_y = m_\psi$) and we see that 
for $\epsilon \sim 1$, and $\phi_0$ determined by the leading (quartic)
higher dimensional gravitational contribution to the superpotential,  
Eq. (\ref{eenonc}) results in
the observed baryon asymmetry: ${n_B}/{s} \sim 10^{-10}~$. So the leading
effects of quantum gravity in the minimal supersymmetric standard  
model act to
generate baryon number violation, and potential terms limiting the flat
direction $vev$'s, of precisely the magnitude needed to generate the  
cosmological 
baryon asymmetry from flat direction oscillations! This indicates how  
robust
the mechanism is, and demonstrates that even in the absence of  
electroweak
baryogenesis, the minimal supersymmetric standard model (including  
the effects of gravity) is
capable of generating the BAU of the right order of magnitude. We also
note that since gravity will   violate $B-L$
(provided we do not gauge $B-L$; we examine GUT extensions of the gauge  
group
below) one expects it to induce operators that violate $B-L$ (eg.  
through $L$
violating operators acting on flat directions), and the resulting $B-L$  
and hence
$B$ asymmetry will be immune to sphaleron erasure.

\section{Regulation by GUT-scale Interactions}

We now turn to a consideration of the effects on flat direction
oscillations of the presence of a new large scale in addition to the  
Planck scale; in particular, we examine baryogenesis  
from flat direction oscillations in the context of Grand Unified Theories. 
Once   one embeds the standard model in a larger gauge theory, at a high (but
sub-Planckian) scale $M_{X}$, we have the possibility of both new  
sources of baryon number violation, and new possible contributions to the  
dynamics of the flat directions.
In general, for all GUT structures from the minimal SU(5) gauge group on up,
there will be induced baryon and lepton number violating effects in low 
energy processes, including baryon- and lepton-number violation inducing quartic 
contributions to the potential for flat direction oscillations (after 
supersymmetry breaking) \cite{ad}.

On the other hand, a generic flat direction of the low energy supersymmetric
standard model may, or may not, have its flatness lifted by GUT interactions.
For those flat directions that are GUT-flat, we expect that
the $vev$'s will ultimately be 
limited by the quantum gravity effects discussed above.
If we assume that the GUT flat direction is lifted by the leading higher
dimensional superpotential term (quartic) which could be induced by 
nonperturbative gravitational effects, then as seen above this yields a 
flat-direction $vev$ after inflation of order  $O(10^{-4})M_{P}$, which
when  substituted into the expression for the resulting GUT induced 
baryon asymmetry from flat
direction oscillations (Eq. (\ref{eenonb})), yields, for $\epsilon$ of order one, 
a baryon
to entropy ratio of order $10^{-4}$. If for some reason the leading 
gravitational superpotential operator vanished, the higher dimensional 
operators would stabilize the flat direction $vev$ at an even larger value 
during inflation, resulting in an even more severe surplus of baryons.

This problematic overproduction of baryons will be regulated if the flat
direction responsible for baryogenesis is flat within the supersymmetric 
standard model, but has its flatness lifted by GUT interactions. 
However, it is important to note that GUT-scale $D$-term interactions can
never lift  flat directions in the absence of supersymmetry breaking effects
as  long as they are $F$-flat.  This is because of the theorem that one 
can always perform a complexified gauge transformation on an $F$-flat 
configuration to make it also $D$-flat \cite{LT}.
This is counter intuitive since one may think that a flat direction can be 
lifted if it is no longer $D$-flat above the 
GUT-scale because of the enhanced gauge symmetry.  Consider, for 
instance, the $L = H_{u}$ flat direction \cite{my} which is $D$-flat 
under the full standard model gauge group.  Since all $D$-flat 
directions in supersymmetric gauge theories are characterized by 
gauge-invariant polynomials \cite{LT}, it is convenient to use the 
language of the polynomials.  The flat direction under discussion 
corresponds to the gauge-invariant $L H_{u}$.  If the gauge group is 
enhanced to $SO(10)$ where $L$ belongs to a {\bf 16} and $H_{u}$ 
to a {\bf 10}, this polynomial is no longer gauge invariant.  
Therefore this flat direction {\it stops}\/ at the GUT-scale.  
However, there must be one or more Higgs fields which break the GUT
gauge  group to the standard model gauge group, and a certain combination of 
the Higgs field and the standard model flat direction remains 
flat.  This is because one can always write a GUT-invariant 
polynomial using powers of the Higgs fields on top of the standard 
model gauge-invariant polynomial.  This 
can be seen explicitly by studying the potential including the 
$U(1)_{X}$ $D$-term together with the GUT-Higgs field $\chi$, 
$\bar{\chi}$ (here they are
assumed to belong to ${\bf 126}$ and $\overline{\bf 126}$) with a 
superpotential $W = (\bar{\chi} \chi - v^{2}) S$:
\begin{equation}
	V = |\bar{\chi} \chi - v^{2}|^{2}
	+ |S|^{2} (|\chi|^{2} + |\bar{\chi}|^{2})
	+ \frac{g^{2}}{8} (|H_{u}|^{2} - |L|^{2})^{2}
	+ \frac{g^{\prime 2}}{8} ( 2 |H_{u}|^{2} - 3 |L|^{2}
		+ 10 |\chi|^{2} - 10 |\bar{\chi}|^{2})^{2} ,
\end{equation}
where $g$ is the $SU(2)$ coupling and $g' = g/\sqrt{10}$ is the 
$U(1)_{X}$ coupling and $S$ is an SO(10) singlet. 
$U(1)_X$ is a subgroup of SO(10) which is not contained in SU(5). One can see
that along the direction
$\chi = v  e^{\eta}$, $\bar{\chi} = v e^{-\eta}$, $|L| = |H_{u}| = l$ and 
$l^{2} = 20 v^{2} \sinh 2 \eta$, the potential remains flat.  This is 
because that one can construct a gauge-invariant polynomial $({\bf 
16}\ {\bf 10})^{10} {\bf 126}$, and hence according to the theorem, 
there is a $D$-flat direction.

Therefore, new $F$-term interactions to the GUT-scale fields are necessary 
to lift flat directions in the standard model.  For the above case, 
it can be the superpotential coupling ${\bf 16}\ {\bf 16}\ {\bf 
126}$ which makes the right-handed neutrino of the ${\bf 16}$ 
heavy.  The net effect then 
is the GUT-scale suppressed effective superpotential such as $(L 
H_{u})(L H_{u})/M_{X}$.  

In general
specific GUTs may have both GUT-flat and GUT-non-flat directions which are 
mutually exclusive (we describe a toy model as an example below). By mutually
exclusive, we mean that the GUT-flat direction is 
no longer flat once the GUT-non-flat direction turns on.
As indicated above, the GUT-flat direction, with GUT scale baryon number
violation leaves us with too large an asymmetry (O($10^{-4}$).  In contrast,
a GUT-non-flat direction, produces a mush smaller flat direction $vev$
given by eq. (\ref{cut}) with the replacement $M_P \to M_X$.  For $M_X \sim
10^{3}$ and $\phi_0^4 \sim 10^{-22}$, (\ref{eenonb}) gives a baryon asymmetry
of the desired magnitude.  Thus, given a GUT with mutually exclusive GUT-flat 
and GUT-non-flat directions, unless one can a priori determine which if any
is preferred, $n_B/s$ is undetermined. In this case, a very mild form of the
anthropic principle is needed: our existence indicates that a GUT-non-flat
direction was chosen (perhaps randomly) over the GUT-flat one.
In fact, mutual exclusivity is probably a desirable feature of the GUT with
respect to the baryon asymmetry.  In its absence, GUT-flat directions would
remain flat despite the population of GUT-non-flat directions.  In this
case, once again, too large an asymmetry would be produced along the GUT-flat
directions.  Thus all GUT-flat directions must be mutually exclusive of some
GUT-non-flat direction.

To illustrate such an example, consider a toy model with a ``GUT'' 
gauge group $U(1)\times U(1)'$, with gauge couplings $g=g'$,
that is broken to $U(1)$ at the scale $M_G$ by $vev$'s of scalar fields
with only $U(1)'$ charge.  We introduce the light superfields
$\Phi_1,\Phi_2,\Phi_3,\Phi_4$ with $U(1)\times U(1)'$ charges 
$(1,1),(-1,1),(-1,-1),(2,0)$, respectively, and superpotential
$$ W_L = \lambda \Phi_4\Phi_2\Phi_3. $$
(Anomaly cancellation in both the low energy and GUT theories is assumed to be 
achieved by additional $U(1)$-charged fields with no flat directions.)
If $\lambda \ne 0$ there are two mutually exclusive flat directions of the low
energy theory, defined by ($\phi_i$ is the scalar component of 
$\Phi_i$) $|\phi_1| = \phi,\; |\phi_4| = 0,$
and  $a$) $|\phi_2| = \phi,\;|\phi_3| =0,$
 or $b$) $|\phi_3| = \phi,\;|\phi_2| =0.$  If $\lambda=0$ ({\it e.g.,} if there 
is no field $\Phi_4$), there is a one parameter family of flat directions 
defined by $\phi = \phi_1 = \phi_2/\sin\alpha  = \phi_3/\cos\alpha $, which 
connects $a$) and $b$).  Above the GUT scale we introduce the massive
superfields $\Pi_0,\Pi_1,\Pi_2,\Pi_3$ with $U(1)\times U(1)'$ charges 
$(0,0),(0,1),(0,-1),(0,-2)$, respectively, and superpotential
$$ W_L = \lambda_0\Pi_0\(\Pi_2\Pi_1-v^2\) + \lambda_3\Pi_1\Pi_3^2 +
\eta\Pi_3\Phi_2\Phi_1. $$ 
(We could also add the gauge invariant term $\zeta\Pi_0\Phi_3\Phi_1$ but this 
does not affect the discussion; it just shifts the $vev\;<\pi_1\pi_2>$ when
$\phi\ne 0$ along the flat direction $b$).)  The only GUT flat direction is
$b$).  If $\lambda=0$, the GUT-scale couplings can drive $\alpha\to0$ without
stabilizing $\phi$.  However if $\lambda\ne0$, and the the light fields lie 
along the flat direction $a$), $\phi$ can be stabilized by the GUT 
interactions. The effective cut-off for the low energy theory is $\Lambda =
M_G$, but loops from the massive fields have the effect of shifting $\Lambda =
M_G\to \Lambda = M_P$ in (\ref{finalV}).  However, integrating out these heavy
fields yields higher dimension operators in the effective low energy tree 
potential for $\phi$ along $a$):
\beq V_H = {\eta^2\(\eta^2 - 2g^2\)\phi^6\over2\lambda_3v^2 +
\(\eta^2 - 2g^2\)\phi^2} -{2g^4\phi^8\over2\lambda_0^2v^4 - g^2\phi^4} =
{\eta^2\(\eta^2 - 2g^2\)\phi^6\over2\lambda_3v^2} + O\(\phi^2\over v^2\).
\label{lep}\eeq
Assuming $\lambda_0^2\sim1,\;0<\eta^2\(\eta^2 - 2g^2\)\sim1,\;
<V>\ll M_G$, we obtain a value during inflation:
$\phi \sim V^{1\over4}M_G \sim 10^{-5.5}$ for $H\sim 10^{-7}, M_G\sim 10^{-2}$.

\section{Partial Sphaleron Erasure}

Another possible suppression mechanism of the baryon asymmetry is to 
employ the smallness of the fermion masses.  The baryon asymmetry is 
known to be wiped out if the net $B-L$ asymmetry vanishes because of 
the sphaleron transitions at high temperature (sphaleron erasure).  
However, Kuzmin, Rubakov and Shaposhnikov (KRS) \cite{KRS} pointed out that
this  erasure can be partially circumvented if the individual $(B-3L_{i})$ 
asymmetries, where $i=1,2,3$ refers to three generations, do not 
vanish even when the total asymmetry vanishes.  Even though there is 
still a tendency that the baryon asymmetry is erased by the
chemical equilibrium due to the sphaleron transitions, the finite mass 
of the tau lepton shifts the chemical equilibrium between $B$ and 
$L_{3}$ towards the $B$ side and leaves a finite asymmetry in the 
end.  Their estimate is
\begin{equation}
	B = - \frac{4}{13} \sum_{i} \left(L_{i} - \frac{1}{3}B\right)
		\left( 1 + \frac{1}{\pi^{2}} \frac{m_{l_{i}}^{2}}{T^{2}}\right)
\end{equation}
where the temperature $T \sim T_C \sim 200$~GeV is when the sphaleron 
transition freezes out (similar to the temperature of the electroweak phase
transition)  and 
$m_{\tau}(T)$ is expected to be somewhat smaller than $m_{\tau}(0) = 
1.777$~GeV. Overall, the sphaleron transition suppresses the baryon 
asymmetry by a factor of $\sim 10^{-6}$.  This suppression factor is
sufficient to keep the total baryon asymmetry at a reasonable order of
magnitude in many of the cases discussed above.

Such $B-3L_{i}$ asymmetries between different generations can be 
generated by the operator
\begin{equation}
	\tilde{m}^{2} \frac{L_{i}^{*} L_{j} H_{u}^{*} H_{u}}{M_{X}^{2}} ,
\end{equation}
which violates individual $B-3L_{i}$ symmetries but not the total 
$B-L$ symmetry. Indeed, the generation of different lepton flavor asymmetries
is expected within the Affleck-Dine framework.
One should note however, that the KRS suppression is only viable if 
lepton number violating interactions for all three generations remain
out-of-equilibrium.  If  one or more, but not all generations have lepton
number violating interactions in equilibrium, then once again a large baryon
asymmetry will result \cite{cdeo}.  If all generations have interactions in
equilibrium, then the asymmetry will be totally washed away by sphaleron
effects.

\section{Conclusions}

We have considered several possibilities for controlling the potentially
large baryon asymmetry produced by the Affleck-Dine mechanism in supergravity
models with a Heisenberg symmetry of the K\"ahler potential.  Moduli, which
normally produce a problematic amount of entropy were shown in fact to be
insufficient in diluting the baryon asymmetry in models with $R$-parity
conservation.  In the absence of $R$-parity, the untwisted sector of orbifold
compactifications may provide a candidate for the desired dilution. 
Perhaps the simplest and most natural possibility for the reasonable baryon
asymmetry invokes only gravitational interactions for both the baryon number
violation as well as the higher dimensional operators necessary for
controlling the sfermion $vev$s along flat directions.  In a GUT, we showed
that it is necessary to have mutually exclusive GUT-flat and GUT-non-flat
directions.  GUT-flat directions do not provide ample suppression of the 
baryon asymmetry.  Finally, we noted that the mechanism suggested by KRS
using lepton mass effects via sphaleron interactions could dilute an
asymmetry by a factor of $10^{-6}$.  

Before concluding we note that string theory predicts a relation between the
values of the gauge coupling $g$ at unification and the scale of unification,
$\Lambda_G
\sim g M_P$, that is  not satisfied in fits of the data to the MSSM.
It has recently been realized that in the strongly coupled limit of heterotic
superstring compactifications (``M-theory"), string unification may in fact be
effectively 5-dimensional \cite{w,p}, with the extra dimension ``turning
on" at $O(10^{14})$ GeV for the gravitational modes of the theory.  In this 
scenario the two $E_8$'s of the heterotic string live on two 
ten-dimensional surfaces that are separated by this fifth (or eleventh) 
dimension $r$. This changes the dimensional analysis that relates the Plank 
scale to the string tension, and modifies the above relation in such a way that
consistency with MSSM RGE's can be achieved for the choice $r\sim 10^4$.
One might expect that $r^{-1}\sim 10^{-4}$ would represent an upper limit for
values of 
$vev$'s that may be reliably considered in the low energy effective theory. 
However, just as the MSSM gauge coupling RGE's are left intact, because the 
observed sector is contained in a (weakly coupled) $E_8$ that does not see the 
extra dimension, the AD flat directions that are charged under the MSSM
gauge group will also be unaffected. Unless effective interactions at the
scale of the fifth dimension can be generated along the flat directions
in such a way so as to preserve the running of the gauge couplings, these
effects will not be capable of reducing the baryon asymmetry.

\vskip 0.8truecm
\noindent {\bf Acknowledgments}
\vskip 0.4truecm
This work was supported in part by the Director, Office of Energy
Research, Office of High Energy and Nuclear Physics, Division of High
Energy Physics of the U.S. Department of Energy under Contracts
DE-FG02-94ER-40823 and 
DE-AC03-76SF00098, by NSF grants AST-91-20005 and PHY-95-14797 and by the 
Alfred P. Sloan Foundation.

\newpage

\section*{Figure Captions}

{\bf Figure 1:} 
Features of the baryon asymmetry in the $\phi_0-m_y$ plane.
In the upper right hand corner of the plot (above the light grey line), the
baryon asymmetry is
$n_B/s \sim O(1)$ and is due to the late decays of sfermion 
oscillations which have come to dominate the energy density of the 
Universe.  Between the grey solid and dashed lines, the sfermions decay
earlier, while the Universe is dominated by the radiation products
of either $y$ or $\psi$ decay (depending on whether $m_y$ is less than
$m_\psi$ or not. Below the dashed line the sfermion flat directions
decay  before the $y$'s or $\psi$'s. Below the grey line,
the baryon asymmetry is given by eq. \protect\ref{ynb} unless
$m_y > m_\psi$ in which case it is given by eq. \protect\ref{eenonb}.
Also shown on the figure (in black) are lines of constant $n_B/s = 10^{-10}$
and
$10^{-11}$.  $m_\psi = 10^{-7}$, with the Hubble parameter during inflation,
$H_I \sim m_\psi$, as well as $\tilde m = 10^{-16}$   were
chosen. In this case, Planck scale baryon number violation ($M_X \sim 1$) was
assumed.  Finally, the shaded region to the left is excluded if R-parity is
conserved.

 \newpage

\begin{figure}[htb]
\vskip 1in
\hskip 1in
\epsfysize=7truein
\epsfbox{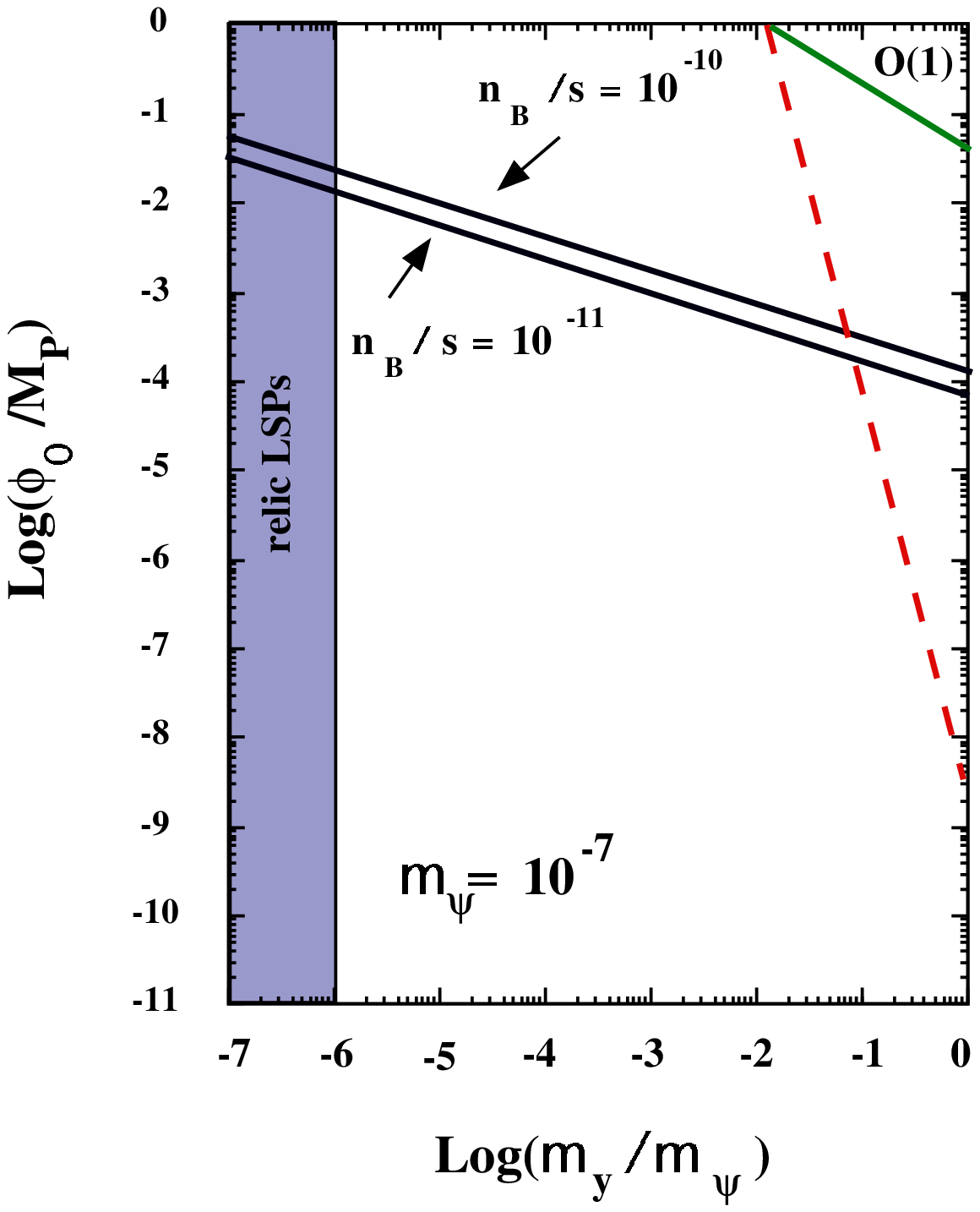}
\end{figure}  

\end{document}